\documentclass[a4paper]{article}
\usepackage{rengo-en}
\usepackage[dvipdfmx]{graphicx}
\usepackage{amsmath}
\usepackage{amssymb}
\usepackage{amsfonts}
\usepackage{color}
\usepackage{graphicx}

\usepackage{unit}
\usepackage{amsmath}
\usepackage{amssymb}
\usepackage{color}
\usepackage{colortbl}

\newcommand{\bm}{\mathbf}
\graphicspath{
{./figs/}
}

\etitle{
Application of Gaussian Process Regression to \\
Koopman Mode Decomposition for Noisy Dynamic Data
\vskip 4mm
}
\eauthor{
Akitoshi Masuda$^{1}$,
Yoshihiko Susuki$^{1,2}$\thanks{Contact information: susuki@eis.osakafu-u.ac.jp, susuki@ieee.org},
Manel Mart\'{i}nez-Ram\'{o}n$^{3}$\\
Andrea Mammoli$^{3}$, 
Atsushi Ishigame$^{1}$
\\
\vspace*{2mm}
$^{1}$Osaka Prefecture University
~~$^{2}$JST, PRESTO
~~$^{3}$The University of New Mexico
}
\englishabstract{
Koopman Mode Decomposition (KMD) is a technique of nonlinear time-series analysis that originates from point spectrum of the Koopman operator defined for an underlying nonlinear dynamical system.
We present a numerical algorithm of KMD based on Gaussian process regression that is capable of handling noisy finite-time data.
The algorithm is applied to short-term swing dynamics of a multi-machine power grid in order to estimate oscillatory modes embedded in the dynamics, and thereby the effectiveness of the algorithm is evaluated.
}

\begin{document}

\maketitle

\section{Introduction}

Koopman Mode Decomposition (KMD) is a novel technique of nonlinear time-series analysis based on spectral properties of the linear but infinite-dimensional composition operator, called the Koopman operator \cite{KMD-Igor,KMD-a}.
The main advantage of KMD is that it is \emph{dynamics-oriented}, implying that it has a solid mathematical foundation in operator theory of nonlinear dynamical systems.
Hence, it has been applied to data-driven method of analysis and control of complex systems (see \cite{Kaiser:2019} and references therein) such as power grids, which include stability analysis \cite{KMD-b} and stabilizing control \cite{KMD-c}.

A numerical method of KMD is generally called the Dynamic Mode Decomposition (DMD) and provides a finite-dimensional approximation of the Koopman operator directly from finite time-series data: see \cite{DMD-a,DMD-b,Kaiser:2019}.
Many variants of the DMD are reported. 
Arnoldi-type (Companion-based) \cite{KMD-a} and Prony-based DMD \cite{Susuki:CDC15} use the idea of Krylov subspace to fit observed current data, which is spanned by past time-series data.
The so-called Extended DMD (EDMD) was proposed by William et al. \cite{EDMD-a} as a generalization of the standard DMD.  
In EDMD, we utilize a finite-dimensional space spanned by finite linearly-independent functions, on which the Koopman operator acts, and approximate its action through a technique of linear regression based on available time-series data.
An application of kernel method to EDMD was also proposed by William et al. \cite{EDMD-b}, and its formulation in Reproducing Kernel Hilbert Space (RKHS) was proposed by Kawahara \cite{KMD-RKHS} and Fujii and Kawahara \cite{KMD-vvRKHS}.

In this report, following \cite{KMD-GP,KMD-GP2}, we address the application of Gaussian Process (GP) regression \cite{GP,Multi-task} to Arnoldi-type DMD. 
GP regression is a powerful approach to Bayesian machine learning and is a method using a probabilistic model to predict output from input.
In \cite{KMD-GP,KMD-GP2}, we used the GP regression to estimate a finite-dimensional approximation of the action of a Koopman operator directly from time-series data, which is an extension of EDMD.
This type of extension is also reported in \cite{KMD-GP3}.
In this report, we use the multi-task GP regression \cite{Multi-task} to derive a new variant of Arnoldi-type (Companion-type) DMD.
A multi-task GP is a framework of multi-task learning in the content of GPs and is capable of handing multiple related tasks.
In connection with KMD, vector-valued time-series from a nonlinear system can be considered to be outputs of latent functions that follow probability functions.
This will provide a robust approach to the projection of noisy data onto the Krylov subspace, which is inevitable in real-world applications such as the mode estimation of power grids directly from data \cite{KMD-b}.
In this report, we first present a formulation for conducting the GP regression directly from time-series data to compute the Koopman eigenvalues and Koopman modes. 
We then apply the GP regression-based algorithm to short-term swing dynamics in the New England 39-bus test grid (NE grid).
The NE grid is a widely-used benchmark for transient stability studies of multi-machine power grids \cite{NE-system} and exhibits coupled swings of the ten synchronous generators operating onto it \cite{Susuki_JNS,KMD-b,KMD-c}, which we refer to as the coupled swing dynamics.

The rest of this report is organized as follows:
in Section~\ref{sec:KMD} we provide the brief introduction to KMD.
In Section~\ref{sec:Computation}, we propose a numerical algorithm for KMD using GP regression.
In Section~\ref{sec:Application}, we apply the proposed algorithm to simulation data of the NE grid with observation noise.

\section{Introduction to Koopman Mode Decomposition}
\label{sec:KMD}

In this section, based on \cite{KMD-Igor,KMD-a,Marko}, we introduce the Koopman operator for nonlinear dynamical systems. 
KMD is a nonlinear time-series analysis based on point spectrum of the Koopman operator.
Now, consider the following finite-dimensional, discrete-time dynamical system: for discrete time $k\in\mathbb{Z}$ and state $\bm{x}\in \mathbb{R}^m$,
\begin{equation}
\bm{x}_{k+1} = \bm{F}(\bm{x}_{k})
\label{eq:DynSyst}
\end{equation}
where $\bm{F}:\mathbb{R}^m\to\mathbb{R}^m$ is a nonlinear continuous map.
To introduce the Koopman operator, we here consider the so-called observable $f:\mathbb{R}^m\to\mathbb{C}$ as a scalar-valued function defined on the state space $\mathbb{R}^m$.
Below, we will denote by $\mathcal{F}$ a (Banach) space of observables.
The Koopman operator ${U}:\mathcal{F}\to\mathcal{F}$ is then defined as a map of $f\in\mathcal{F}$ into a new function by
\begin{equation}
{U}f := f\circ\bm{F}.
\end{equation}
An important point of it is that even if the original system \eqref{eq:DynSyst} is nonlinear, the Koopman operator is linear.
Thus, our idea is to investigate dynamics described by the nonlinear system \eqref{eq:DynSyst} through the linear operator $U$.

Here, we introduce the KMD. 
To do this, consider a vector-valued (multi-task) observable $\bm{f}={[f_1,\ldots,f_M]^{\top}}:\mathbb{R}^m\to\mathbb{C}^{M} (f_i\in\mathcal{F})$ where ${\top}$ stands for the transpose operation.
Let $\psi_{j}\in\mathcal{F}\setminus\{0\}$ be the $j$-th eigenfunction of ${U}$ with associated Koopman Eigenvalue (KE) $\lambda_{j}\in\mathbb{C}$:
\begin{equation}
{U}\psi_{j}=\lambda_{j}\psi_{j}.
\label{eq:Def.}
\end{equation}
The KE exists for a wide class of nonlinear systems, and the cardinality of all KEs can be countably infinite \cite{Marko}.
If each observable $f_i$ lies in the subspace spanned by all the eigenfunctions, then as in \cite{KMD-a}, it can be expanded as follows:
\begin{equation}
f_i=\sum_{j=1}^{\infty}\psi_{j}v_{ij}
\label{eq:EigenExpa.}
\end{equation}
where $v_{ij}$ are complex-valued constants for the expansion.
Using (\ref{eq:DynSyst}) and \eqref{eq:Def.}, the time evolution $\bm{f}(\bm{x}_k)$ is expanded as
\begin{equation}
\bm{f}(\bm{x}_k)
=\underline{{U}}^k\bm{f}(\bm{x}_0)=\sum_{j=1}^{\infty}\lambda_{j}^{k}\psi_{j}(\bm{x}_0)\bm{v}_{j}
\label{eq:KMD}
\end{equation}
where $\underline{{U}}^k\bm{f}:={[{U}^kf_1,\ldots,{U}^kf_M]^{\top}}$ and $\bm{v}_j:=[v_{1j},\ldots,v_{Mj}]^\top\in\mathbb{C}^M$.
The expansion (\ref{eq:KMD}) implies that the multi-task observables $\bm{y}_k=\bm{f}(\bm{x}_k)$ of (\ref{eq:DynSyst}) is decomposed into an infinite sum of modes oscillating with single frequencies.
KE $\lambda_{j}\in\mathbb{C}$ characterizes the frequency and Growth Rate (GR) of each mode.
The constant vector $\bm{v}_{j}\in\mathbb{C}^{M}$, called Koopman Mode (KM), represents the modal contribution to every task in the component of $\bm{y}_k$.
This type of time-series analysis is coined by Rowley et al. \cite{KMD-a} as the KMD (Koopman Mode Decomposition).

\section{Application of Gaussian Process Regression to Koopman Mode Decomposition}
\label{sec:Computation}

This section presents the main contribution of this report: we provide a numerical algorithm of KMD based on GP regression in order to KEs and KMs directly from time-series data.

Consider a finite-length sequence of $N+1$ multi-task observations of \eqref{eq:DynSyst} as $\left\lbrace\bm{y}_{0},\bm{y}_{1},\dots,\bm{y}_{N}\right\rbrace$  where $\bm{y}_k\in\mathbb{R}^M$.
For the application of GP regression, we need input/output data as a training dataset.
It is here recalled that GP regression has been used in the context of dynamical modeling: see, e.g., \cite{GP-time-series,GP-time-series2}.
Following this, in order to handle the dynamics, we suppose that the $k$-th snapshot $\bm{y}_k$ as output is determined by the $p$ past snapshots  $\{\bm{y}_{k-p},\ldots,\bm{y}_{k-1}\}$ as input, where $p$ is a positive constant.
We denote the input by $\bm{z}_{k}:=[\bm{y}_{k-p}^{\top},\ldots,\bm{y}_{k-1}^{\top}]^{\top}\in\mathbb{R}^{M\cdot p}$.
Then, $N-p+1$ training samples are available for the current GP regression:
\[
\left\{(\bm{z}_{p},\bm{y}_{p}),(\bm{z}_{p+1},\bm{y}_{p+1}),\dots,(\bm{z}_{N},\bm{y}_{N})\right\}.
\]

Here, we use the formulation of multi-task GP regression in \cite{Multi-task}.
We consider latent functions $\bm{g}={[g_1,\ldots,g_M]^{\top}}:\mathbb{R}^{M\cdot p}\to\mathbb{R}^{M}$
which derive a latent distribution $\bm{g}(\bm{z}_k)$ in terms of real (noisy) output $\bm{y}_k$.
We also consider a GP prior over the latent functions $\bm{g}$ and assume that GPs possess zero mean and covariance given by
\begin{equation}
\text{cov}(g_i(\bm{z}_{k}),g_j(\bm{z}_{l})) := [{\bf K}^g]_{ij}\kappa(\bm{z}_{k},\bm{z}_{l})
\end{equation}
where $\kappa: \mathbb{R}^{M\cdot p}\times\mathbb{R}^{M\cdot p}\to\mathbb{R}$ is a covariance (kernel) function over input $\bm{z}$ and ${\bf K}^g\in\mathbb{R}^{M\times M}$ is a positive semi-definite matrix specifying inter-task similarities.
By assuming additive independent Gaussian noise with variance $\sigma_i^2$ for each task $i$, the prior distribution of the observations (multi-task outputs) possesses the following covariance: 
\begin{equation}
\text{cov}([\bm{y}_{k}]_i,[\bm{y}_{l}]_j) = [{\bf K}^g]_{ij}\kappa(\bm{z}_{k},\bm{z}_{l})+\sigma_i^2\delta_{ij}\delta_{kl}
\end{equation}
where $[\bm{y}_{k}]_i$ stands for the $i$-th element of vector $\bm{y}_{k}$ and $\delta_{ij}$ is the Kronecker delta.

We define the complete set of training outputs for $M$ tasks as $\bm{y}:=[\bm{y}_{p}^{\top},\ldots,\bm{y}_{N}^{\top}]^{\top}\in\mathbb{R}^{M\cdot (N-p+1)}$ and the GP predictive values $\bm{g}(\bm{z}_{N+1})$, which are associated with the multi-task observables of \eqref{eq:DynSyst}.
By assuming that both the outputs and predictive values obey the following prior distributions, they are described as follows:
\begin{align}
&\begin{bmatrix}
\bm{y}\\
\bm{g}(\bm{z}_{N+1})
\end{bmatrix}\sim \mathcal{N}\left({\bm 0},\right.\nonumber \\
&
\left.
\begin{bmatrix}
{{\bf K}(\bm {z},\bm {z})}\otimes{\bf K}^g+{\bf I}\otimes{\bf D} & {{\boldsymbol \kappa}(\bm {z},\bm{z}_{N+1})}\otimes{\bf K}^g\\
{\boldsymbol \kappa}(\bm{z}_{N+1},\bm {z})\otimes{\bf K}^g & {\kappa(\bm{z}_{N+1},\bm{z}_{N+1})}\otimes{\bf K}^g
\end{bmatrix}
\right) 
\end{align}
where $\mathcal{N}$ denotes the normal distribution and $\otimes$ the Kronecker product.
$\bm {z}$ is the set of $N-p+1$ inputs $\bm{z}_{p},\dots,\bm{z}_{N}$,
${\bf K}(\bm {z},\bm {z})$ is the Gram matrix given by
$(\kappa({\bm z}_k,{\bm z}_l))_{k,l=p,p+1,\dots,N}$,
and ${\boldsymbol \kappa}(\bm {z},\bm{z}_{N+1})$ is the $(N-p+1)\times1$ vector of the covariances evaluated at all pairs of training input and test input $\bm{z}_{N+1}$.
${\bf D}$ is an $M\times{M}$ diagonal matrix in which the $(i,i)$-{th} element corresponds to $\sigma_i^2$.
Then, according to \cite{Multi-task}, the predictive conditional distribution becomes
\begin{align}
\bm{g}(\bm{z}_{N+1})|\bm {z},\bm{y},\bm{z}_{N+1} & \nonumber\\
& \makebox[-2em]{}
\sim \mathcal{N} \left(\overline{\bm{g}(\bm{z}_{N+1})},\text{cov}(\bm{g}(\bm{z}_{N+1}))\right),
\label{eq:mean_predict}
\end{align}
with
\begin{align}
\overline{\bm{g}(\bm{z}_{N+1})}
=& {{\boldsymbol \kappa}(\bm{z}_{N+1},\bm {z})}\otimes{\bf K}^g \nonumber\\
& \times[{{\bf K}(\bm {z},\bm {z})}\otimes{\bf K}^g+{\bf I}\otimes{\bf D}]^{-1}\bm{y}, \\
\text{cov}(\bm{g}(\bm{z}_{N+1}))
=& {\kappa(\bm{z}_{N+1},\bm{z}_{N+1})}\otimes{\bf K}^g \nonumber \\
& -{{\boldsymbol \kappa}(\bm{z}_{N+1},\bm {z})}\otimes{\bf K}^g \nonumber\\
& \times[{{\bf K}(\bm {z},\bm {z})}\otimes{\bf K}^g+{\bf I}\otimes{\bf D}]^{-1} \nonumber\\
& \times{{\boldsymbol \kappa}(\bm {z},\bm{z}_{N+1})}\otimes{\bf K}^g.
\end{align}

Finally, we derive a decomposition formula similar to a finite truncation of \eqref{eq:KMD}.
The predictive mean values derived above become
\begin{equation}
\overline{\bm{g}(\bm{z}_{N+1})} = {\bf B}{\boldsymbol \kappa}(\bm {z},\bm{z}_{N+1}),
\label{eq:predict_mean}
\end{equation}
with
\begin{equation}
{\bf B} := {\bf K}^g{\bf H}
\end{equation}
where ${\bf H}\in\mathbb{R}^{M\times(N-p+1)}$ and
\begin{equation}
\text{vec}({\bf H}):= [{\bf K}(\bm {z},\bm {z})\otimes{\bf K}^g+{\bf I}\otimes{\bf D}]^{-1}\bm{y}.
\end{equation}
The above derivation is based on the following formula:
\begin{equation}
\text{vec}({\bf XYZ}) = ({\bf Z^{\top}}\otimes{\bf X})\text{vec}({\bf Y}),
\label{eq:lemma}
\end{equation}
with
\begin{align}
{\bf X} &= {\bf K}^g,\\
{\bf Y}&={\bf H},\\
{\bf Z}&={\boldsymbol \kappa}(\bm {z},\bm{z}_{N+1}),\\
{\bf Z}^{\top}&={\boldsymbol \kappa}(\bm {z}_{N+1},\bm{z}).
\end{align}

In the same manner, the mean values of latent functions at $\bm{z}_{k}$ are given by
\begin{equation}
\overline{\bm{g}(\bm{z}_{k})} = {\bf B}
{\boldsymbol \kappa}(\bm {z},\bm{z}_{k}),~~~k=p,p+1,\dots,N.
\label{eq:latent_func}
\end{equation}
Here, we define ${\bf G}\sub{GP}$ and $\bm{c}\sub{GP}$ as
\begin{align}
{\bf G}\sub{GP} :=& \left[\overline{\bm{g}(\bm{z}_{p})},\overline{\bm{g}(\bm{z}_{p+1})},\ldots,\overline{\bm{g}(\bm{z}_{N})}\right]\\
=& {\bf B}{\bf K}(\bm {z},\bm {z})\in\mathbb{R}^{M\times(N-p+1)}, \nonumber\\
\label{eq:GPc}
\bm{c}\sub{GP} :=& {\bf K}(\bm {z},\bm {z})^{-1}{\boldsymbol \kappa}(\bm {z},\bm{z}_{N+1}),
\end{align}
If ${\bf K}(\bm {z},\bm {z})$ is not regular, then we use the Moore-Penrose pseudo-inverse for computation.
Using \eqref{eq:predict_mean} and \eqref{eq:latent_func}, we have
\begin{align}
\underline{{U}}{\bf G}\sub{GP}
&= \left[\overline{\bm{g}(\bm{z}_{p+1})},\overline{\bm{g}(\bm{z}_{p+2})},\ldots,\overline{\bm{g}(\bm{z}_{N+1})}\right] \nonumber\\
&=: {\bf G}\sub{GP}{\bf C}\sub{GP}
\label{UGeqGC'}
\end{align}
where ${\bf C}\sub{GP}$ is the $(N-p+1)$-dimensional companion matrix defined as
\begin{equation}
{\bf C}\sub{GP} :=
\begin{bmatrix}
    0 & 0 & \cdots & 0  & c\sub{GP,0}\\
    1 & 0 &  & 0  & c\sub{GP,1}\\
    0 & 1 &  & 0 &  c\sub{GP,2}\\
    \vdots &  & \ddots &   & \vdots\\
    0 & 0 & \cdots & 1 & c\sub{GP,{\it N-p}+1}
\end{bmatrix}.
\label{defC}
\end{equation}
The ${\bf G}\sub{GP}$ contains the mean values of latent functions that are derived by removing noise from the training outputs.
The $\bm{c}\sub{GP}$ is considered to be the coefficient vector of linear regression of the predictive mean's values by the mean values of latent functions.
Then, we locate the $N-p+1$ eigenvalues of ${\bf C}\sub{GP}$, called the \emph{Ritz} values $\tilde{\lambda}_j$ ($j=1,2,\ldots,N-p+1$), and we define the Vandermonde matrix ${\bf T}\sub{GP}$ as follows:
\begin{equation}
	{\bf T}\sub{GP} :=
	\begin{bmatrix}
			1 & \tilde{\lambda}_{1} & \tilde{\lambda}_{1}^{2} & \cdots & \tilde{\lambda}_{1}^{N-p}\\
			1 & \tilde{\lambda}_{2} & \tilde{\lambda}_{2}^{2} & \cdots & \tilde{\lambda}_{2}^{N-p}\\
			\vdots & \vdots & \vdots & \ddots & \vdots \\
			1 & \tilde{\lambda}_{N-p+1} & \tilde{\lambda}_{N-p+1}^{2} & \cdots & \tilde{\lambda}_{N-p+1}^{N-p}
		\end{bmatrix}.
\label{defT}
\end{equation}
Here, the \emph{Ritz} vectors $ \tilde{\bm{v}}_j$ are defined to be the columns of ${\bf V}\sub{GP}:={\bf G}\sub{GP}{\bf T}\sub{GP}^{-1}.$
By assuming that the $\lambda_j$ are distinct, then the following expansion of the output at time $k+p$ holds: 
\begin{align}
{\bm{y}}_{k+p} =& \sum_{j=1}^{N-p+1}\tilde{\lambda}_{j}^{k}\tilde{\bm{v}}_{j}+\bm{r}_k,~~~
\nonumber \\
& \makebox[4em]{}k=0,1,\ldots,N-p+1
\label{eq:GPDMD}
\end{align}
where $\bm{r}_k := {\bm{y}}_{k+p}-\overline{{\bm{g}}(\bm{z}_{k+p})}$ corresponds to the error vector of GP regression due to the mean values of latent functions.

\section{Mode Estimation of Multi-Machine\\ Power Grid}
\label{sec:Application}

\begin{figure}[tb]
\centering
\includegraphics[width=0.49\textwidth]{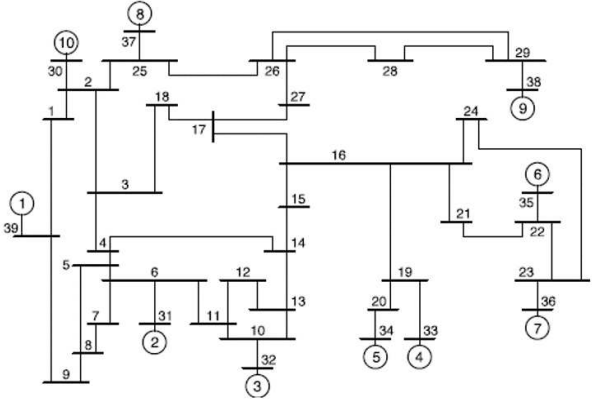}
\caption{New England 39-bus test grid (NE grid)}
\label{fig:NE-39bus}
\end{figure}

We apply the GP regression-based algorithm to analyze short-term swing dynamics exhibited in the New England 39-bus test grid (NE grid).
The NE grid is shown in Fig.\,\ref{fig:NE-39bus} and contains the ten generation units (equivalent ten synchronous generators, circled numbers in the figure), the 39 buses, and AC transmission lines.
Most of the buses have constant active and reactive power loads.
The details of the system, such as unit rating, line data, and loading conditions, are given in \cite{NE-system}.

\subsection{Nonlinear Swing Equations}
\label{subsec:Swing Equations}

First, we introduce the equations of motion of generators in the NE grid.
Assume that bus 39 is the infinite bus \cite{PSSC}.
The short-term electro-mechanical dynamics of generators 2--10 are represented by the following nonlinear differential equations, called the classical model \cite{PSSC}:
\begin{equation}\label{eq:Classical-Model}
\left.
\begin{aligned}
\frac{\dd\delta_i}{\dd t}
&=\Delta\omega_i\\
\frac{H_i}{\pi{f_b}}\frac{\dd\Delta\omega_i}{\dd t}
&=P_{{\rm m}i}-D_i\Delta\omega_i-P_{{\rm e}i}\makebox[1em]{}
\end{aligned}
\right\},
\end{equation}
with
\[
P_{{\rm e}i}
=\sum_{j=1}^{10}E_iE_j\{G_{ij}\text{cos}(\delta_i-\delta_j)+B_{ij}\text{sin}(\delta_i-\delta_j)\}
\]
where the integer label $i=2,\ldots,10$ denotes generator $i$.
The variable $\delta_i$ is the angular position of rotor in generator $i$ with respect to bus 1 and is in radians [rad].
The variable $\Delta\omega_i$ is the deviation of rotor speed in generator relative to that of bus 1 and is in radians per second [rad/s].
We set the variable $\delta_1$ to a constant, because bus 39 is assumed to be the infinite bus. 
The parameters $P_{{\rm m}i}, E_i, G_{ij}$ and $B_{ij}$ are in per unit system, $H_i$ and $D_i$ are in seconds [s], and $f\sub{b}$ is in Hertz [Hz].
The mechanical input power $P_{{\rm m}i}$ to generator $i$ and the internal voltage $E_i$ of generator $i$ are normally constant in the short-term regime \cite{PSSC}.
The parameter $H_i$ is the per unit time inertia constant of generator $i$, and $D_i$ its damping coefficient.
The parameter $G_{ii}$ is the internal conductance, and $G_{ij}+\text{j}B_{ij}$ is the transfer admittance of the $(i,j)$-th element of the reduced admittance matrix of the grid.
Any electrical loads are modeled as passive impedances.
Note that any model of exciter and controller is not included in the model.

\subsection{Numerical Simulation}
\label{subsec:Simulation}

The setting of numerical simulation is based on \cite{Susuki:2011}.
The voltages $E_i$ at a stable equilibrium ($\delta_i^\ast,\Delta\omega_i^\ast=0$) for generator $i$ are fixed using power flow computation. 
The constants $H_i$ are the same as in \cite{NE-system}, $P_{{\rm m}i}$ and power loads are half of the rating in \cite{NE-system}.
The parameters $D_i$ are fixed at $0.05\U{s}$, and $f\sub{b}$ at $60\U{Hz}$.
The elements $G_{ij}$ and $B_{ij}$ are calculated using the data in \cite{NE-system} and the power flow computation.
All numerical simulations were performed using MATLAB: the function \texttt{ode45} was adopted for numerical integration of (\ref{eq:Classical-Model}).

\begin{figure}[tb]
\centering
\includegraphics[width=0.52\textwidth]{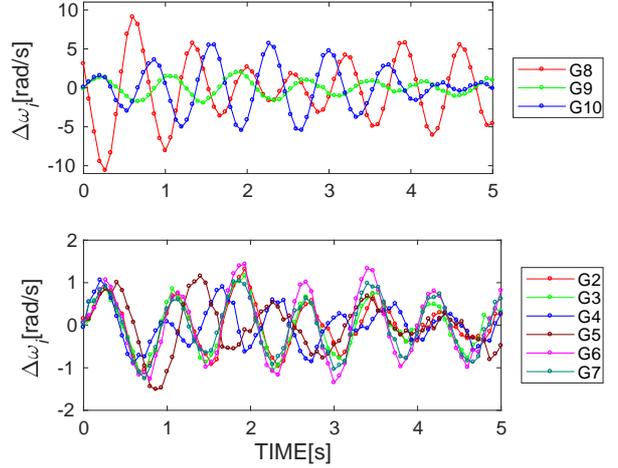}
\caption{%
Rotor speed deviations of generators 2--10 in the New England test grid.
These are the trajectories of (\ref{eq:Classical-Model}) for the initial condition (\ref{eq:initial condition}) and contain additive i.i.d. noise.
}
\label{fig:data}
\end{figure}

We present an example of short-term dynamics in the NE grid. 
Fig.\,\ref{fig:data} shows the time responses of rotor speed deviations $\Delta\omega_i$ under the initial condition from \cite{Susuki:2011}:
\begin{equation}\label{eq:initial condition}
(\delta_i(0),\Delta\omega_i(0))=
\begin{cases}
(\delta_i^\ast+1.5~\text{rad},3~\text{rad/s}) &i=8,\\
(\delta_i^\ast,0~\text{rad/s}) &\text{else}.
\end{cases}
\end{equation}
The simulation contains i.i.d. noise $\mathcal{N}\left(\bm{0},0.1^2\right)$ for each time-series
as an observation noise, which aims to evaluate the effectiveness of the GP-based algorithm.
The initial condition physically corresponds to a local disturbance at generator~8.
The generators do not show any stepping-out in the figure, that is, they do not show any loss of transient stability for the selected disturbance.
Generators~8 and 10 have swings of larger amplitudes than the others,
because the initial condition is localized at generator 8, and the
two generators are electrically close.
Even for the additive noise, the swing dynamics of generators observed here are the same as in \cite{Susuki:2011}.

\subsection{Computation of Koopman Modes and Eigenvalues}
\label{subsec:Koopman-arnoldi}

Next, we compute the KEs and KMs (empirical Ritz values $\tilde{\lambda}_j$ and vectors $ \tilde{\bm{v}}_j$) for the coupled swing dynamics shown in Fig.\,\ref{fig:data}. 
The computation is investigated in the two different manners.  
One is to show a representative result of applying GP-based algorithm to the time-series data.
The other is to assess noise dependency of the result.
For the computation, we need to choose the observable $\bm{f}(\bm{\delta},\bm{\Delta\omega})$, where $\bm{\delta}:=[\delta_2,\dots,\delta_{10}]^{\top}$ and $\bm{\Delta\omega}:=[\Delta\omega_2,\dots,\Delta\omega_{10}]^{\top}$.
In this application, we use the rotor speed deviations $\bm{\Delta\omega}$:
\[
\bm{f}(\bm{\delta},\bm{\Delta\omega})=\bm{\Delta\omega}.
\]
The sampling period $T$ of the simulation output for the application is $1/(15\U{Hz})$.
The analysis window is fixed at $[0\U{s},4\U{s}]$, and thus the number of samples corresponds to $N=60$.

\subsubsection{Representative result}

\begin{figure}[t]
\centering
\includegraphics[width=0.52\textwidth]{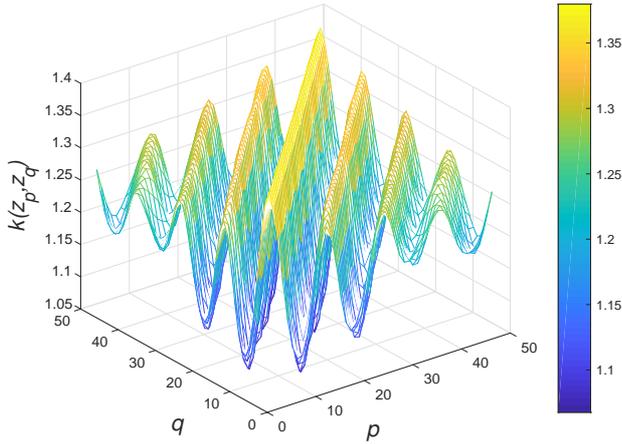}
\caption{%
Values of the Gaussian kernel  $\kappa(\bm{z}_p,\bm{z}_q)$ for the set of input data sampled from Fig.\,\ref{fig:data}
}
\label{fig:kernel_data}
\end{figure}

First, we apply the GP-based algorithm to the simulation output in Fig.\,\ref{fig:data}.
As training inputs of GP, we take the past observations with $p=15$, \emph{i.e.}, $\bm{z}_{k}={[\bm{y}_{k-15}^{\top},\ldots,\bm{y}_{k-1}^{\top}]^{\top}}\in\mathbb{R}^{135}$.
We also take $\bm{y}_k\in\mathbb{R}^{9}$ as the training output.
The kernel function used in this report is the well-known Gaussian kernel \cite{GP} given by
\begin{equation}
\kappa(\bm{z}_p,\bm{z}_q)
:=\sigma\sub{f}^2\exp\left\{-\frac{1}{2\ell^2}
\left(\sum_{i=1}^{135}\left|\frac{[\bm{z}_p]_i}{s_i}
-\frac{[\bm{z}_q]_i}{s_i}\right|^2\right)\right\}
\end{equation}
where the parameters $\sigma\sub{f}^2$ and $\ell$ (and noise variance $\sigma\sub{i}^2$, too) are determined by Leave-One-Out-Cross-Validation (LOO-CV) \cite{GP}.
$s_i$ is called the scaling parameter \cite{GP-time-series} corresponding to $\max\{[\bm{z}_k]_i|; k=N_p,N_p+1,\dots,N\}$, and the multi-task outputs $\bm{y}_k$ are also scaled by the parameters $ss_i$ corresponding to $\max\{|[\bm{y}_k]_i|; k=0,1,\dots,N\}$.
Fig.\,\ref{fig:kernel_data} shows the values of Gaussian kernel for the set of input data sampled from Fig.\,\ref{fig:data}.
The values quantify the similarity between different two inputs in terms of the Gaussian kernel, that is, the inner product inside the induced RKHS.
Furthermore, we here set the intertask covariance ${\bf K}^g$ as follows: 
\begin{equation}
{\bf K}^g = \text{diag}
(ss^{-1}_1,ss^{-1}_2,\ldots,ss^{-1}_M) \quad (ss_{i}\neq 0).
\end{equation}
This setting is basically for tractable computation and implies that each of the normalized tasks has the same variance of observation noise. 
\begin{figure}[tb]
\centering
\includegraphics[width=0.53\textwidth]{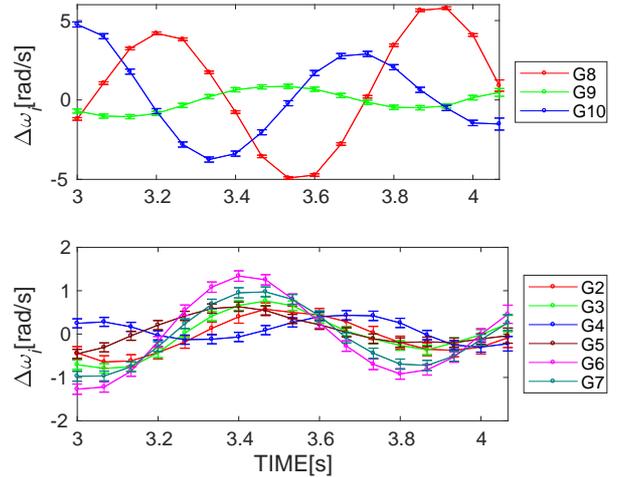}
\caption{%
GP predictive means and associated 95\% confidence intervals
}
\label{fig:predict_data}
\end{figure}
Fig.\,\ref{fig:predict_data} shows the result of GP regression, that is, the GP predictive means and associated 95\% confidence intervals. 
By using the means, we compute the KEs and KMs.
Now let us focus on KMs that have both large growth rates $|\tilde{\lambda}_j|$ and large norms of $\tilde{\bm{v}}_j$.
Such modes represent sustaining components for the time duration of simulation output and have dominant magnitudes in the output. 
Table~\ref{table_Single} shows the numerical result on dominant KEs and KMs for Fig.\,\ref{fig:data}, which we call Mode\,1 to Mode\,7.
The norm for Mode\,$j$ is defined as the standard Euclidean norm $\|\tilde{\bm{v}}_j\|$.
Below, we pick up Mode 1 (period $0.67\U{s}$) and Mode 2 (period $0.79\U{s}$) with largest norms in Table~\ref{table_Single}.  
By observation of Fig.\,\ref{fig:data}, Mode 1 can be confirmed as a dominant swing component in generators 8 and 10, and
Mode 2 can be confirmed in generators 2, 3, 6, 7, and 9.

\begin{table}[tb]
	\centering
	\caption{%
	Computational result on Koopman modes and eigenvalues for Fig.\,\ref{fig:data} 
	}%
	\label{table_Single}
	\vskip 4mm
		\begin{tabular}{cccc}\hline
			Mode            & Norm  & Growth & Period [s] \\
			& & Rate & \\
			$j$      & $\|\tilde{\bm{v}}_j\|$   & $|\tilde{\lambda}_j|$  & $T_j:=2\pi{T}/\text{Im}[\text{ln}\tilde{\lambda}_j]$ 
			\\
			\hline
			\rowcolor[rgb]{0.7, 0.7, 1.0} 1	& 3.136	& 0.995	& 0.671 \\
			\rowcolor[rgb]{1.0, 1.0, 0.7} 2	& 2.662	& 0.994	& 0.788 \\
		  3	& 0.651	& 0.983	& 0.999 \\
		  4	& 0.502	& 0.985	& 1.903 \\
		  5	& 0.088	& 0.976	& 0.352 \\
			6	& 0.078	& 0.979	& 0.389 \\
			7	& 0.043	& 0.989	& 0.220 \\
      \hline
		\end{tabular}
\end{table}

\begin{figure}[t] 
\centering
\includegraphics[width=0.49\textwidth]{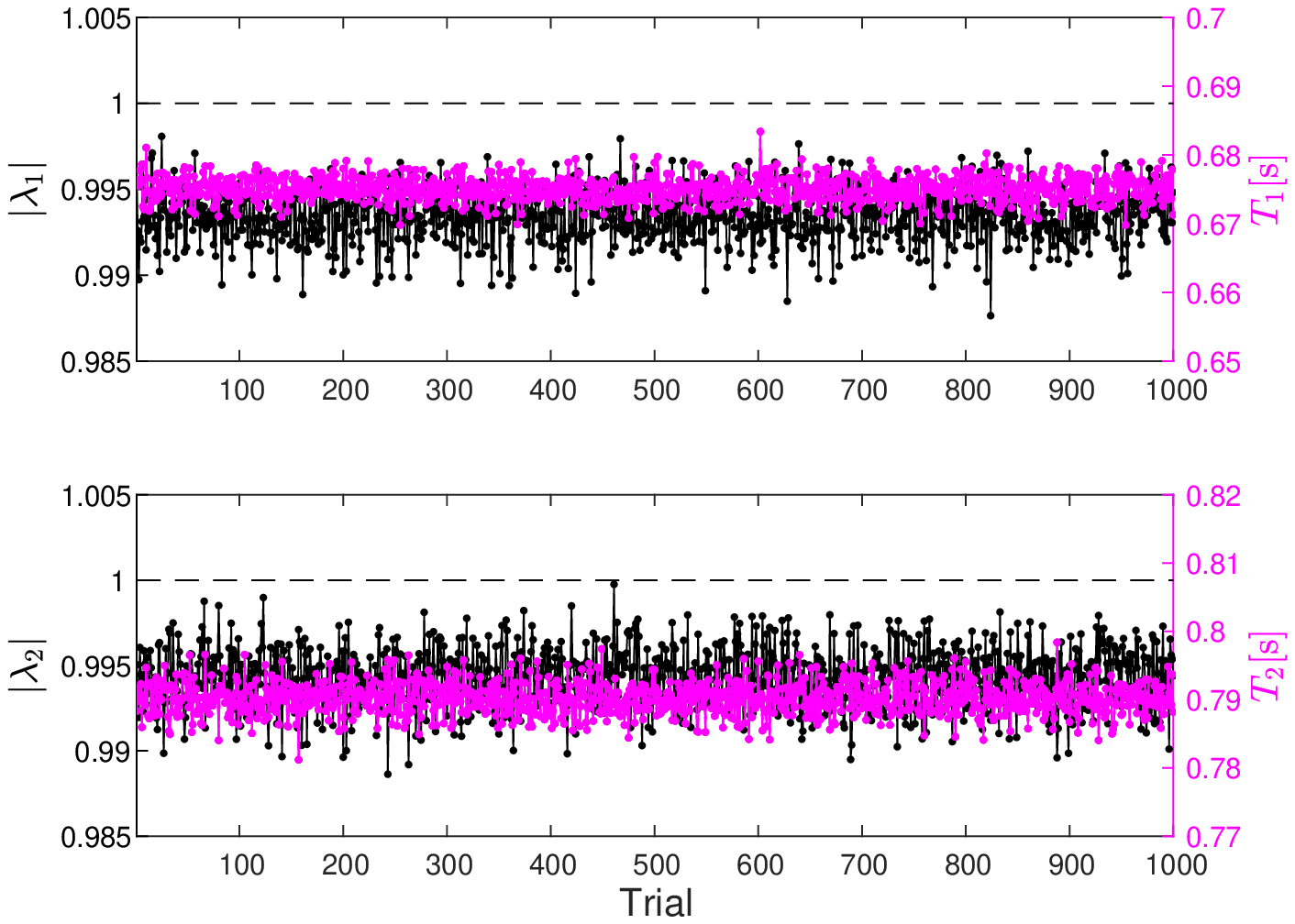}
\vspace*{-6mm}
\caption{%
Noise dependency of the magnitude of Koopman eigenvalues (left; \emph{black}) and of their periods (right; \emph{purple}) 
}
\label{fig:GP-eigen}
\vspace*{3mm}
\centering
\includegraphics[width=0.52\textwidth]{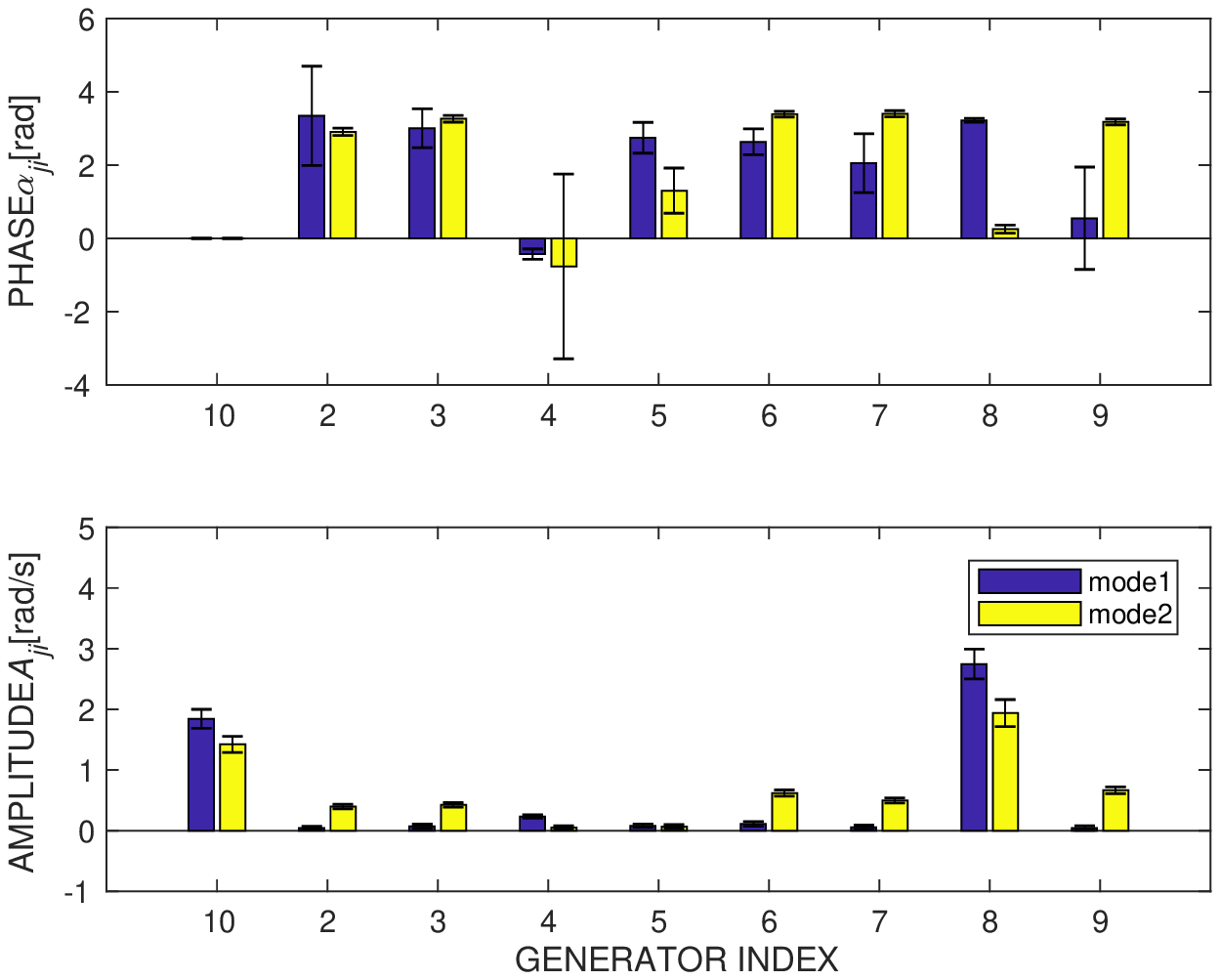}
\vspace*{-6mm}
\caption{
Shapes of the two Koopman modes $\tilde{\bm{v}}_j$ ($j=1,2$) in Table~\ref{table_Single}.
The 95\% confidence intervals are shown with the error bars.
}
\label{fig:GP-mode}
\end{figure}

\subsection{Assessment of noise dependency}

Here, we assess how the computed KE and KM are affected by the additive noise.
In Fig.\,\ref{fig:GP-eigen} we show the computational results on the two modes for 1000 samples of additive noise obeying $\mathcal{N}\left(\bm{0},0.1^2\right)$.
The trial number is labeled on the horizontal axis of the figure.
Thus, by considering the 95\% intervals, we have the following estimation of magnitude of Mode\,1 and Mode\,2, namely GR (Growth Rate), and their periods:
\begin{equation}
\left.
\begin{array}{ccl}
|\tilde{\lambda}_1| 
&\in& [0.9943-0.0034,0.9943+0.0034]\\
T_1 
&\in& [0.6750\,{\rm s}-0.0036\,{\rm s},0.6750\,{\rm s}+0.0036\,{\rm s}]
\end{array}
\right\},
\label{eq:eig1}
\end{equation}
and
\begin{equation}
\left.
\begin{array}{ccl}
|\tilde{\lambda}_2| 
&\in& [0.9935-0.0030,0.9935+0.0030]\\
T_2 
&\in& [0.7902\,{\rm s}-0.0048\,{\rm s},0.7902\,{\rm s}+0.0048\,{\rm s}]
\end{array}
\right\}.
\label{eq:eig2}
\end{equation}
Clearly, the maximum of GR is smaller than unity.
The result on GR suggests that the GP-based algorithm can work for the data-driven stability analysis \cite{KMD-b} for the dominant modes under additive observation noise. 

In addition, we investigate the mode shapes for Mode\,1 and Mode\,2 by the amplitudes $A_{ji}$ and the phases $\alpha_{ji}$ (with respect to generator 10) of each generator: 
\begin{equation}
\left.
\begin{array}{lcl}
A_{ji} &:=& |\tilde{v}_{j,(i+8)}|,~~~i=2,\ldots,10\\
\alpha_{ji} &:=& \text{Im}[\text{ln} (\tilde{v}_{j,(i+8)}/\tilde{v}_{j,18})]
\end{array}
\right\}.
\label{eq:ap}
\end{equation}
In particular, the mean and standard deviation of $A_{ji}$ and $\alpha_{ji}$ ($j=1,2$ and $i=2,\dots,10$) are evaluated for all the results of $1000$ samples of additive noise.
The evaluation is shown in Fig.\,\ref{fig:GP-mode} where the \emph{blue} (or \emph{yellow}) bars are for Mode\,1 (or Mode\,2).
The error bars denote the 95\% confidence intervals in this analysis.
The standard deviations of $\alpha_{1,2}$, $\alpha_{1,7}$, $\alpha_{1,9}$, $\alpha_{2,4}$, and $\alpha_{2,5}$ are very large because the amplitudes $A_{1,2}$, $A_{1,7}$, $A_{1,9}$, $A_{2,4}$, and $A_{2,5}$ are nearly zero, and hence the associated values of phases are not tractable.
The other standard deviations are small, implying that the mode estimation is reliable.
Mode 1 has a small amplitude other than generators 8 and 10, and their phases $\alpha_{1,8}$ and $\alpha_{1,10}$ have opposite phases.
This implies that Mode 1 represents an inter-machine swing mode between generators 8 and 10. 
Also, Mode 2 has a large amplitude other than generators 4 and 5, and their phases $\alpha_{2,8}$ and $\alpha_{2,10}$ are in phase and opposite phase to $\alpha_{2,2}$, $\alpha_{2,3}$, $\alpha_{2,6}$, $\alpha_{2,7}$ and $\alpha_{2,9}$.
Mode 2 is regarded as an inter-area swing mode between the group of generators 8 and 10 and the adjacent group of generators 2, 3, 6, 7, and 9. 
As above, we show that the GP-based algorithm as the mode estimation method for the New England test grid is robust against the additive observation noise.


\section{Conclusion}
In this report, we derived a GP (Gaussian Process) regression-based algorithm of KMD (Koopman Mode Decomposition) for noisy dynamic data.
The use of GP regression is expected to robustify the computation of KEs (Koopman Eigenvalues) and KMs (Koopman Modes) under observation noise in the dynamic data.
We applied the GP-based algorithm to data on nonlinear dynamic simulations of the power grid benchmark model.
This shows that the algorithm is robust for mode estimation of the power grid against observation noise.

Our future works are to clarify a systematic way to identify hyper-parameters from a dynamical system perspective and to reformulate the regression problem in this report for approximating the action of Koopman operator. 

\section*{Acknowledgement}

The authors would like to thank Dr. Satomi Sugaya (The University of New Mexico) for her careful reading of an early version of this report and Prof. Y. Kawahara (Kyushu University) for valuable comments. 
The work presented here is supported in part by JSPS-KAKEN Grant Number 15H03964, Japan, The Specific Support Project of Osaka Prefecture University, 
JST, PRESTO Grant Number JPMJPR1926, Japan, and NSF EPSCoR Cooperative Agreement OIA-1757207 and TEC 2014-52289-R, United States.


\end{document}